\documentclass[galaxies,review,accept,pdftex,moreauthors]{Definitions/mdpi} 

\usepackage{graphicx}

\firstpage{1} 
\pubvolume{1}
\issuenum{1}
\articlenumber{0}
\pubyear{2024}
\copyrightyear{2024}
\externaleditor{Academic Editor: }
\datereceived{24 August 2024} 
\daterevised{24 September 2024} 
\dateaccepted{25 September 2024}
\datepublished{ } 
\hreflink{https://doi.org/} 

\newcommand{\ixpe}{IXPE}
\newcommand{\chandra}{\textit{Chandra}}
\newcommand{\rxte}{RXTE}
\newcommand{\sax}{\textit{BeppoSAX}}
\newcommand{\rxj}{RX~J1713.7$-$3946}

\Title{Probing Magnetic Fields in Young Supernova Remnants with~IXPE}

\TitleCitation{Probing Magnetic Fields in Young Supernova Remnants with IXPE}

\Author{Patrick Slane
 $^{1,}$*\orcidA{}, Riccardo Ferrazzoli $^{2}$\orcidB{}, Ping
Zhou $^{3}$\orcidC{}  and Jacco Vink $^{4}$\orcidD{}}

\AuthorNames{Patrick Slane, Riccardo Ferrazzoli, Ping Zhou, and Jacco Vink}

\AuthorCitation{Slane, P.; Ferrazzoli, R.; Zhou, P.; Vink, J.}

\address{%

$^{1}$ \quad Center for Astrophysics | Harvard \& Smithsonian, 60
Garden Street, Cambridge, MA 02138, USA \\
$^{2}$ \quad INAF Istituto di Astrofisica e Planetologia Spaziali,
Via del Fosso del Cavaliere 100, 00133 Roma, Italy ;
riccardo.ferrazzoli@inaf.it\\ 
$^{3}$ \quad School of Astronomy and Space Science, Nanjing University, 
Nanjing 210023, People's Republic of China; pingzhou@nju.edu.cn\\ 
$^{4}$ \quad Anton Pannekoek Institute for Astronomy \& GRAPPA, University 
of Amsterdam, Science Park 904, 1098 XH Amsterdam, The Netherlands; 
j.vink@uva.nl}

\corres{Correspondence: slane@cfa.harvard.edu}

\abstract{Synchrotron emission from the shocked regions in supernova remnants
provides, through its polarization, crucial details about the
magnetic field strength and orientation in these regions.  This,
in turn, provides information on particle acceleration in these
shocks.  Due to the rapid losses of the highest-energy relativistic
electrons, X-ray polarization measurements allow for investigations of the
magnetic field to be carried out
very close to the sites of particle acceleration.
Measurements of both the geometry of the field and the levels of
turbulence implied by the observed polarization degree thus provide
unique insights into the conditions leading to efficient particle
acceleration in fast shocks.  The Imaging X-ray Polarimetry Explorer
(IXPE) has carried out observations of multiple young SNRs, including
Cas~A, Tycho, SN~1006, and \rxj.
In each, significant
X-ray polarization detections provide measurements of magnetic field
properties that show some common behavior but also considerable
differences between these SNRs.  Here, we provide a summary of results
from IXPE studies of young SNRs, providing comparisons between the
observed polarization and the physical properties of the remnants
and their environments.}

\keyword{X-rays; polarimetry; supernova remnants; cosmic ray;
particle acceleration; magnetic field turbulence; individual objects:
Cas A, Tycho's SNR, SN 1006, \rxj}

\begin{document}

\section{Introduction}

Over the past two decades, it has become clear that the rapid shocks
in young supernova remnants (SNRs) are capable of accelerating
particles to extremely high energies. Particle acceleration
most likely proceeds through diffusive shock acceleration (DSA),
wherein energetic particles streaming away from the shock scatter
off of {magnetohydrodynamical} waves---either pre-existing in the
ambient plasma or generated by other streaming ions---and return
to the shock for additional acceleration, e.g.,~\citep{MalkovDrury2001}. 
This process builds up a nonthermal population of high-energy
particles, with the maximum energy being limited by radiative losses,
the age of the SNR, or particle escape~\citep{Reynolds2008}. Such
particle acceleration by SNR shocks has long been suggested as a
process by which Galactic cosmic rays are produced, up to energies
approaching the ``knee'' of the cosmic ray spectrum at $\sim $10$^{14}$--10$^{15}$~eV.

The details of this acceleration process are very much still under
investigation. The standard picture for DSA requires strong magnetic
fields and high levels of turbulence in order for the efficient return
of particles to the acceleration site. Particle transport is
generally described in terms of a mean-free path that is a factor
$\eta$ times the particle gyroradius. The maximum acceleration
efficiency for the loss-limited case that applies to electrons in
young SNRs (because of high synchrotron losses) is obtained for
\mbox{$\eta = 1$}, the so-called Bohm limit for which the mean-free path
is the minimum value. The associated photon cutoff energy for
synchrotron emission from electrons in this process is \mbox{$\epsilon_{0}
\approx 1.6 \eta^{-1} (v_{\rm sh}/4000 {\rm\ km\ s}^{-1})^2 {\rm\
keV}$}~\citep{zirakashvili07}. To produce photons up to several keV,
high shock speeds are required, such as those found in very young
SNRs. In addition, small Bohm factors are required, indicating
highly turbulent magnetic fields.


The above description holds in the so-called ``test particle'' limit
where the energy of the accelerated particles is dynamically
unimportant. If the relativistic particle component of the energy
density becomes comparable to that of the thermal component, the
shock acceleration process can become highly nonlinear. The gas
becomes more compressible, which results in higher magnetic fields
and enhanced acceleration.

For most of the known young SNRs, thin rims of nonthermal X-ray emission
surround the remnants directly along their forward shocks. These
observations make it clear that SNRs are capable of accelerating
electrons to very high energies. The observed rim widths are typically
3--5~arcsec---resolvable only with \chandra---and lead to field
strengths ranging  from 
 $\sim$100 to 500~$\upmu$G, demonstrating significant
amplification of the typical  3~$\upmu$G upstream field in the {interstellar medium}.

The nature of the turbulence that plays such a crucial role in the
acceleration process is not yet fully understood. The cosmic ray
current associated with the highest-energy particles, which escape
the acceleration region, produces turbulence in the upstream region~\citep{Bell2004}. This is ultimately swept up by the shock, where
the component of the field that is tangent
to the shock front is
compressed. This leads to an expectation of synchrotron radiation
with a radial polarization (i.e., with the electric vector oriented
along the radial direction). Such polarization is indeed observed
in radio observations of older SNRs ~\citep{dickel76,dubner15} 
(see the latter reference for a
recent review). However, radio observations of
young SNRs like Cas A, Tycho, and others show tangential polarization,
indicating radial magnetic fields in the downstream region
\citep{dickel76}. The nature of these radial fields is not well
understood. Particle-in-cell simulations indicate that the
acceleration process is more efficient for shocks whose velocity
vectors are parallel to the magnetic field. One suggestion for the
observed parallel fields in young remnants is simply that one
observes radiation from regions where the acceleration is most
efficient~\citep{West+2017}. Alternatively, density fluctuations---possibly in the form of clumps from material upstream---can
form magneto-driven instabilities in the downstream region that
drive radial magnetic fields~\citep{Inoue+2013}, suggesting that
the ambient medium into which the SNRs expand might play a significant
role. Recent studies indicate that the Bell instability can produce
density fluctuations sufficient to drive this same mechanism without
any requirement for a pre-existing clumpy environment~\citep{Bykov+2024}.

In any such mechanism, the question of magnetic field orientation
and the amplitude of the associated turbulence are fundamentally
important. Both of these parameters can be probed with polarization.
Because of the short synchrotron lifetimes of X-ray-emitting
electrons, X-ray polarization explores the magnetic orientation and
turbulence very close to the particle acceleration site, and thus
represents a particularly important tool for constraining the
acceleration process.

In its first two years, the IXPE has observed six SNRs. Here, we summarize
results of the initial four studies, with observations of
Cassiopeia A, Tycho's SNR, the northeast limb of SN 1006, and the
northwest region of \rxj. We present a general introduction
of the properties of each remnant, summarize the IXPE observations
and results for each, and then discuss the results in the context
of models for polarization in SNRs.

\section{Cas A}

Cas~A is the relic of a core-collapse supernova estimated to have
occurred roughly 350 years ago. Located at a distance of 3.4~kpc
\citep{Reed+1995}, it is one of the brightest SNRs in the radio and
X-ray bands. Expansion measurements establish a forward shock
velocity of $\sim$5000${\rm\ km\ s}^{-1}$,
e.g., \citep{Vink+1998,PatnaudeFesen2009}.  Radio polarization
measurements establish a polarization degree (PD) of $\sim$5\% in
the bright ejecta-rich shell and 8--10\% in the outer regions near
the forward shock~\citep{Anderson+1995}. The polarization angle
(PA) implies a radial magnetic field~\citep{Rosenberg1970,
BraunGull1987}.

X-ray observations with \sax\ and \rxte\ revealed the presence of
synchrotron emission from particles with energies up to $\sim $10~TeV
\citep{Favata+1997,Allen+1997}, consistent with the acceleration of
particles by the rapid shocks. \chandra\ observations establish
that the nonthermal X-rays arise from thin regions
\citep{Hughes+2000,Hwang+2004} whose $1^{\prime\prime}$--2$^{\prime
\prime}$ widths imply magnetic field strengths of 250--500~$\upmu$G,
e.g., \citep{VinkLaming2003,Bamba+2005}, implying significant
magnetic field amplification.

\begin{figure}[H]
\includegraphics[width=.98\textwidth]{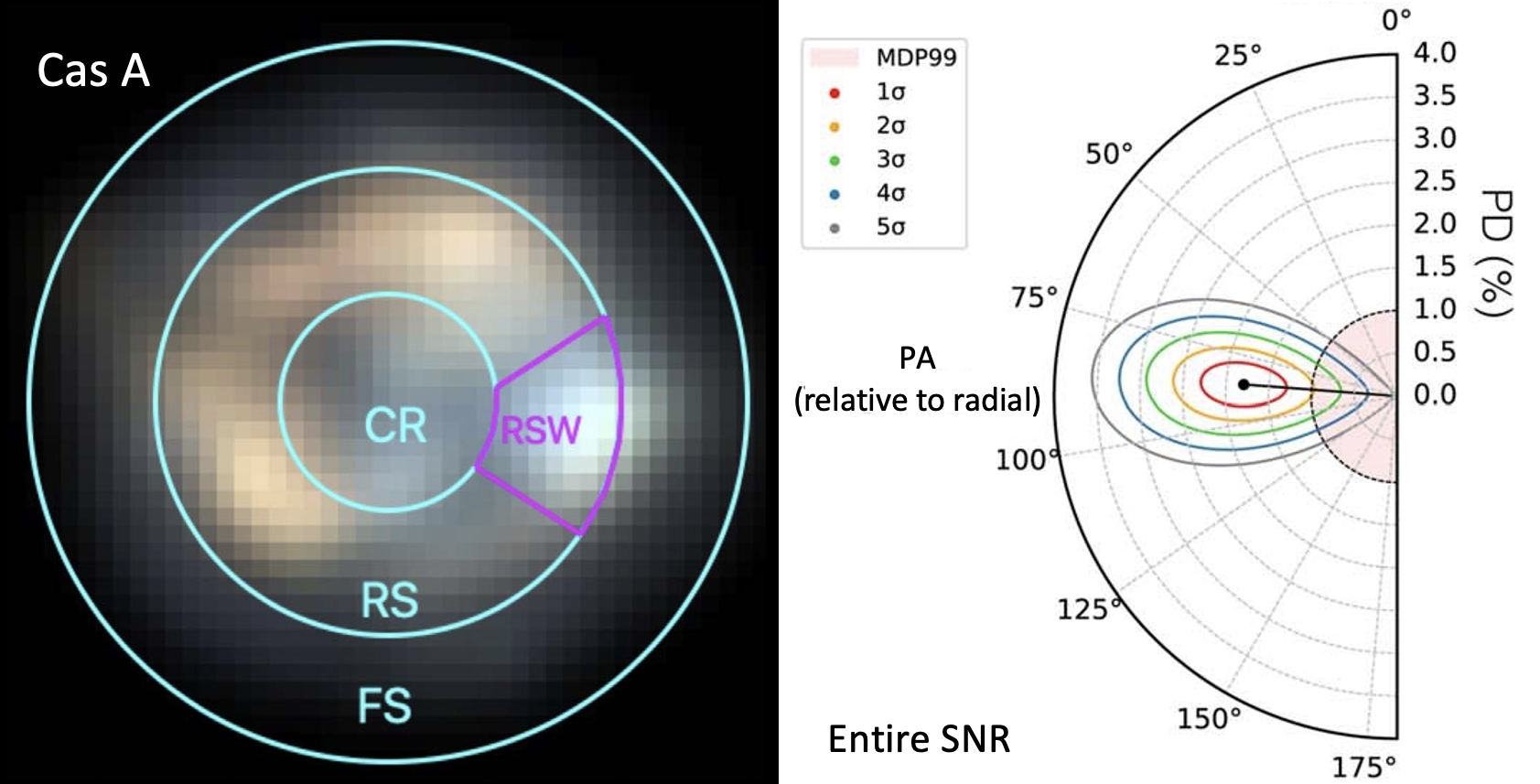}
\caption{\textbf{Left}: IXPE image of Cas A. Labels identify {the central
region (CR), reverse shock (RS), western reverse shock (RSW), and
forward shock (FS)} investigated in~\citep{Vink+2022}. \textbf{Right}: Polar
plot for emission from entire SNR.  The polarization degree is
indicated on the radial axis and the polarization angle is shown,
with $0^\circ$ corresponding to the radial direction; {an angle of
$90^\circ$ indicates a radial magnetic field direction}.  The
contours show significance levels.}
\label{CasA-fig}
\end{figure}

The IXPE observed Cas~A in 2022 for $\sim$900~ks~\citep{Vink+2022}.~A pixel-by-pixel search revealed modest evidence for X-ray polarization
with polarization degrees of 5--15\%, and with indications of
radial magnetic fields. By assuming circular symmetry of the
polarization direction, however, larger emission regions could be
combined to improve sensitivity. These measurements established
highly significant detections of polarization from multiple regions,
including the forward shock and reverse shock, and marginal evidence
for similar polarization from a distinct region on the western edge
of the reverse shock, illustrated in Figure \ref{CasA-fig} (left),
where we present a three-band color IXPE image of Cas~A with these
regions identified. Figure \ref{CasA-fig} (right) shows the
polarization results for the entire SNR, where the polarization
degree is represented by the radial component of the polar plot and
the polarization angle is measured relative to the radial direction.
Contours show confidence intervals, and the pink circle illustrates
the minimum detectable polarization at 99\% confidence for this
observation. {Significant polarization is detected from the forward
shock and reverse shock (RS) regions, along with the entire SNR.}
The PA values for all regions are consistent with
tangential polarization---and thus radial magnetic fields---with
PD values ranging from $\sim$2 to 4\%. Using {\sl Chandra} spectra
for the individual regions to determine fluxes for both the nonthermal
emission and the (unpolarized) thermal emission, the corrected PD
values for the X-ray synchrotron radiation range from $\sim$2 to 5\%---lower than observed in the radio band.

The low polarization degree measured with the IXPE implies high levels of
turbulence in the region close to the shock, implying that despite
the shock compression that should produce a primarily tangential field
at the shock, magnetic fields are reoriented on spatial scales
corresponding to the thin rims ($\sim$10$^{17}$~cm).

\section{Tycho's SNR}

Tycho's SNR is the remnant of the historical supernova SN 1572
\citep{2003Green}.  Unlike Cas A, a core-collapse remnant, Tycho
is the result of a Type Ia explosion.  Its shock velocity is in the
range of 3500--4400 km s$^{-1}$~\citep{2017Williams, 2020Williams}.
Radio emission from Tycho at 2.8--6 cm is polarized, with typical
polarization degree values ranging from 0 at the center to 7--8\%
at the outer rim, with the direction of the polarization indicating
a large-scale radial magnetic field structure~\citep{1971Kundu,1973Strom,
1975Duin, 1991Dickel, 1997Reynoso}.  High-resolution Chandra X-ray
observations of Tycho revealed a bright thin synchrotron rim at the
shock~\citep{2002Hwang} matching similar features seen in the radio
band~\citep{1991Dickel}, but also peculiar small-scale structures
(from $\sim$arcseconds to $\sim$arcminute) in the 4--6\,keV band---the so-called ``stripes'' in the western rim that were first
identified by \cite{2011Eriksen}.  These small-scale structures are
variable in time and shape on a scale of a few years
\citep{2020Okuno,2020Matsuda}.  The stripes, along with the entire
SNR rim, are thought to be CR acceleration sites, and might be the
result of fast energy losses of the TeV electrons emitting X-rays
downstream of the shock in amplified magnetic fields~\citep{2020Bykov}.
If that is the case, the synchrotron structures are due to geometric
projection of the thin regions where the TeV electrons are accelerated; the expectations of their polarization properties were
extensively discussed prior to the IXPE launch
\citep{2009Bykov,2011Bykov_a,2020Bykov}. 


The IXPE observed Tycho during 2022, for a total exposure time of
$\sim$990 ks~\citep{2023Ferrazzoli}. The analysis followed a
strategy similar to the one used for Cas A, starting with a
pixel-by-pixel search of the signal. However, since Tycho is not as
bright as Cas A, the polarization map binned on a 1-arcminute scale
does not show highly significant detections.

\begin{figure}[H]
\includegraphics[width=.98\textwidth]{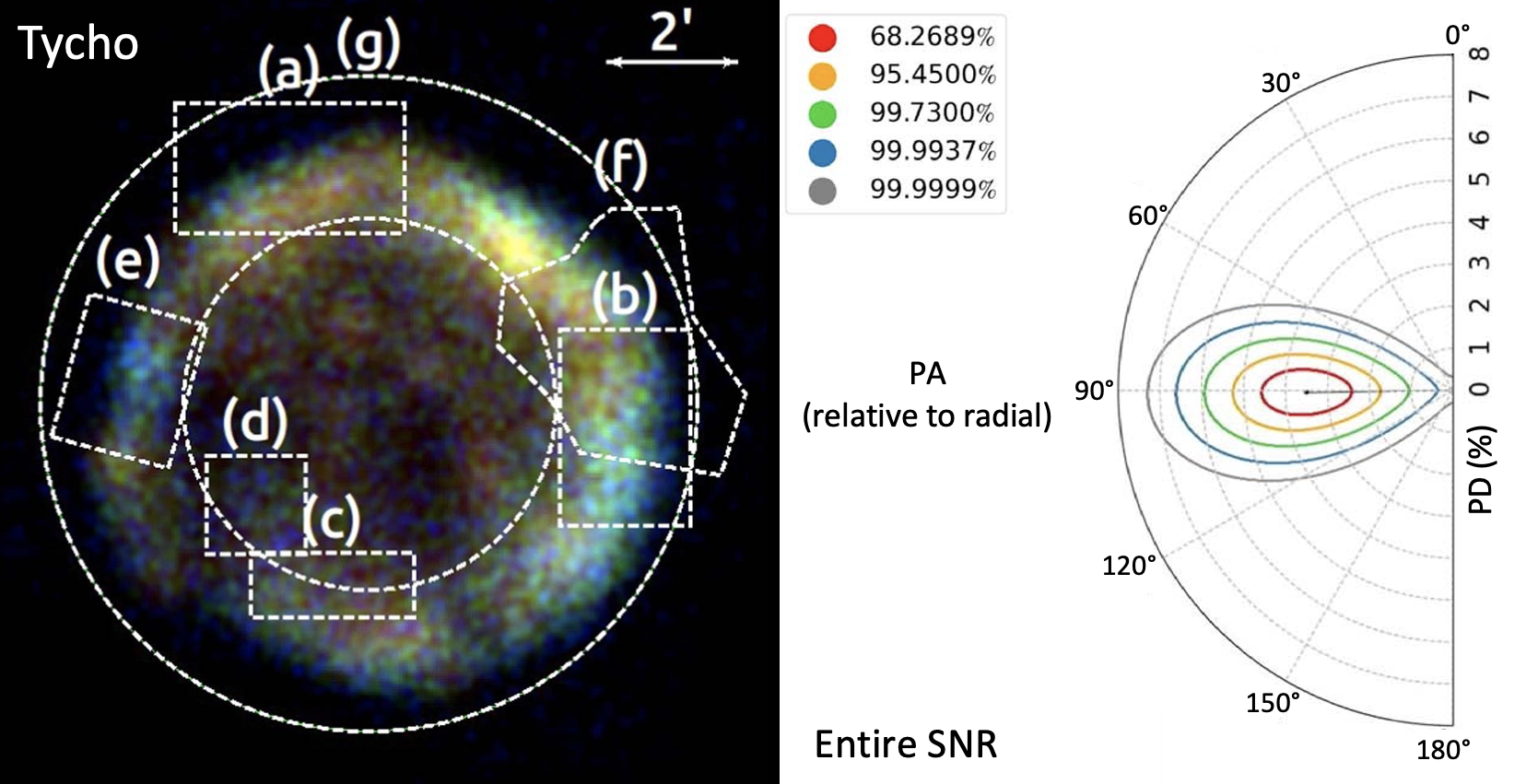}
\caption{\textbf{Left}: IXPE image of Tycho. Labels identify {specific} regions
investigated in~\citep{2023Ferrazzoli}{---the northeast knot (a);
the western (b) and southern (c) nonthermal stripes and the nonthermal
arc region (d); and regions in which strong polarization is detected
(e,f)}. \textbf{Right}: Polar plot for entire SNR {(region g)}.  The
{polarization degree is indicated on the radial axis and the}
polarization angle is shown, with $0^\circ$ corresponding to the
{radial direction; an angle of $90^\circ$ indicates a radial}
magnetic field direction. {The
contours show significance levels.}}
\label{Tycho-fig}
\end{figure}

By aligning and summing up the data from different regions of
interest {shown in Figure \ref{Tycho-fig}}, such as the highest
significance region in the west {(region f)}, the rim {(region g)}, and
also the whole remnant, highly significant detections of polarized
emission were identified.  The measured tangential direction of
polarization corresponds to a radial magnetic field, consistent
with the radio band polarization observations but originating from
regions even closer to the shock.  The degree of X-ray polarization
in Tycho is significantly higher than that for Cas A: in Tycho, the
synchrotron PD was found to be $12\%\pm2\%$ in the rim and $9\%\pm2\%$
in the whole remnant. These results are compatible with the
expectation of turbulence produced by an anisotropic cascade of a
radial magnetic field near the shock.


\section{SN 1006}

SN~1006 is the first SNR detected to show strong nonthermal X-ray
emission~\citep{1995Koyama}, making it a good target for X-ray
polarimetry. The IXPE observed the northeastern shell of SN~1006
(SN~1006~NE) on 5--19 August  2022 
 and  3--10 March 2023, with a total
effective exposure of $\sim$960~ks~\citep{2023Zhou}. Compared to
Cas~A and Tycho's SNR, that of SN~1006 is much fainter in X-rays. The
background contributes a non-negligible fraction of the photons
from the source region, and thus we removed this part in the analysis.

\begin{figure}[H]
\includegraphics[width=.98\textwidth]{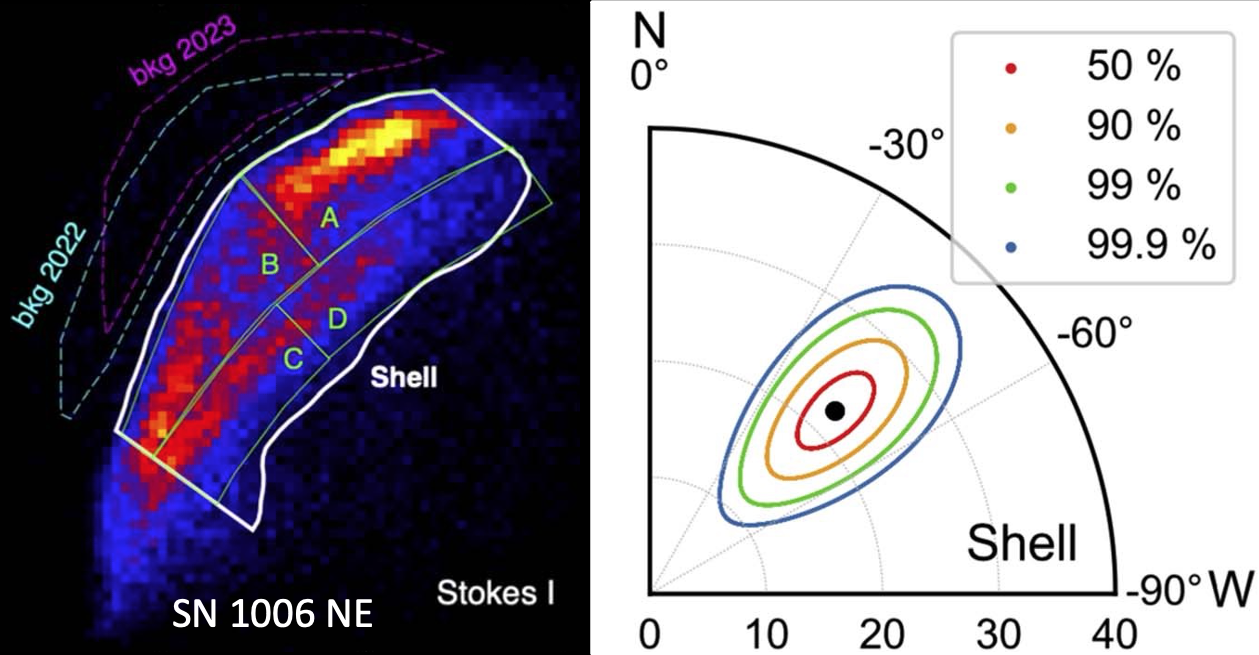}
\caption{\textbf{Left}: IXPE 
 image of the NE limb of SN~1006. Labels identify {discrete
portions of the shell }
investigated in \cite{2023Zhou}. \textbf{Right}: Polar plot for entire shell
region. The polarization angle is shown with $0^\circ$ corresponding to
North. {An angle of $\sim$$-$45$^\circ$ corresponds to tangential polarization,
and thus a radial magnetic field. The
contours show significance levels.}}
\label{fig:sn1006}
\end{figure}

The IXPE resolved the double-rim structure of SN~1006~NE and allowed
for a spatial analysis of the X-ray polarization. The IXPE Stokes
I image of SN~1006~NE in 2--4~keV is shown in Figure~\ref{fig:sn1006},
where the white and green polygon regions denote the shell region
and four sub-scale regions, respectively.  The X-ray polarization
of the overall shell was detected at a significance of $6.3\sigma$,
with a polarization degree of $22.4\pm 3.5\%$ and a polarization
angle of $-45.5^\circ$. The sub-scale regions A--C all show similar
polarization properties with a significance between 3.4 and 4.6,
while region D is marginally detected ($2.5\sigma$). These results
suggest that the polarization degree does not significantly vary
across the NE shell and magnetic fields are nearly radially
distributed.  We also used a different \mbox{approach---spectropolarimetric}
analysis---to calculate the polarization of these five regions and obtained consistent results.

SN~1006 has been measured in radio polarimetry~\citep{2013Reynoso}.
Assuming a constant rotation measure, ${\rm RM=12~rad~m^{-2}}$, we
calculated the radio polarization results for SN~1006~NE and compared
them with the IXPE values.  We found a larger polarization angle
of $-36.3^\circ\pm 0.4^\circ$ and a lower polarization degree of
$14.5\%\pm 0.2\%$ in the radio band. Since the radio results,
especially the polarization angle, were influenced by the $\rm RM$
distribution, which was not well constrained in previous radio
observations, we did not expand the discussion about the discrepancy
between the two bands. However, a lower polarization degree in the
radio band is not unexpected. First, the maximum polarization degree
depends on the photon index, $\Gamma$: $PD_{\rm max}=\Gamma/(\Gamma+2/3)$.
The X-ray spectrum is steeper than the radio band, suggesting a
larger $PD_{\rm max}$.  Secondly, due to the shorter lifetime of
the high-energy electrons, the synchrotron X-ray emission is from
a thinner area than that of the radio emission, meaning that the
radio emission is likely subject to stronger depolarization.

\section{\rxj}

\rxj\ is a large ($\approx$1 degree diameter) shell-type SNR residing
in the Galactic plane, discovered by the ROSAT all-sky survey
\citep{1996Pfeffermann}. The remnant is at a distance of $\sim$1~kpc~\citep{2003Fukui,2003Uchiyama,2004Cassam-Chenai} and is thought
to be the result of a Type Ib/c supernova explosion, possibly
associated to the historical SN 393~\citep{1997Wang,2016Tsuji,2017Acero},
making it the oldest SNR whose results have been reported by the IXPE
so far. Its hard X-ray emission is purely nonthermal
\citep{1997Koyama,1999Slane}, representing the second detection of
synchrotron X-ray radiation from an SNR shell after SN 1006. Tsuji
et al.~\citep{2016Tsuji} measured the shock velocity in the
northwestern region to be $\sim$3900 km s$^{-1}$, with other
structures within the NW shell, however, being significantly slower,
down to $\sim$1400  km s$^{-1}$. Previous attempts at measuring
radio polarization led to detections at the northwestern part of
the shell, although Faraday rotation made it impossible to determine
the magnetic field direction from the polarization vectors
\citep{2004Lazendic}. \rxj\ is also one of the best studied young
SNRs in the $\gamma$-ray band and one of the most debated sources
in the discussion on the hadronic vs. leptonic origin of the high
energy emission.


The IXPE observed the northwestern part of the shell of \rxj\ three
times in 2023, 24--27 August, 28 August to 1 September, and  24 September to 5 October, for a total exposure time of
$\sim$841~ks~\citep{2024Ferrazzoli}.

\begin{figure}[H]
\includegraphics[width=.98\textwidth]{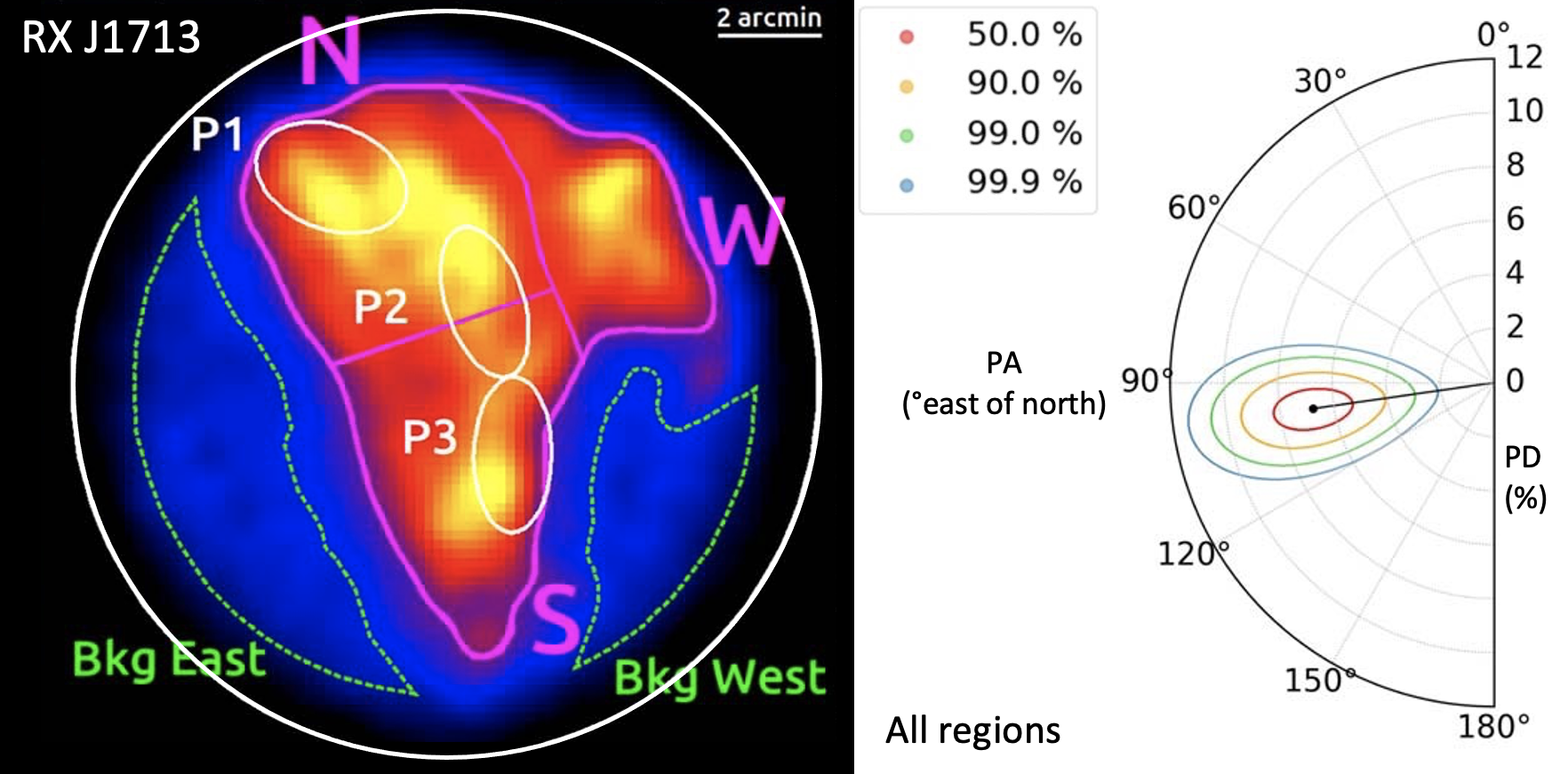}
\caption{\textbf{Left}: IXPE 
 image of the NW limb of \rxj. Labels identify
{north, west, and south (N, W, S) regions of the extended emission
structure and discrete regions P1--P3 that show evidence for higher
polarization---all} were investigated in~\citep{2024Ferrazzoli}. \textbf{Right}:
Polar plot for entire limb region, composed of N + W + S. The polarization
angle is {measured from north to east, with an angle of $\sim$100$^\circ$}
corresponding to the radial direction {from the center of \rxj\ and,
thus, to a magnetic field that is tangential to the shock surface. The
contours show significance levels.}}
\label{RXJ1713-fig}
\end{figure}

In the left panel of Figure \ref{RXJ1713-fig}, the regions of interest
selected for the analysis and the background extraction regions are
shown. {Significant polarization is found for all identified regions
other than the broad western region (W). For the N and S regions, the
polarization degree is modest ($\sim$12--18\%), while for the 
individual regions P1--P3, identified with high significance in binned maps, 
the values are as high as $\sim$45\%}. From both the polarization
map and the regions of interest, our analysis found that the
polarization direction is normal to the shock and that the average
PD in the whole region is {\mbox{$13.0\pm3.5\%$}}. Thus, whereas the other
IXPE remnants all showed radial magnetic fields, \rxj\ represents
the first example for which shock-compressed tangential magnetic
fields dominate in the X-ray band. These results are consistent
with a model of shock compression of an upstream isotropic turbulence
producing a predominantly tangential magnetic field.


\section{Discussion}

The polarization results from \ixpe\ observations of young SNRs
show that the {polarization degree} is quite small (Table \ref{tab1}), indicating
high levels of turbulence in the immediate post-shock regions, as
expected from models for diffusive shock acceleration with magnetic
field amplification. It is noteworthy that the level of {circumstellar
medium (CSM)} interaction---indicated by the ambient density, $n_0$---is high in Cas A (the product of a core-collapse SN expanding
into a CSM highly modified by the progenitor star), modest in Tycho
(but not zero; despite being the remnant of a Type Ia SN, Tycho
shows evidence of CSM interactions), and small in SN~1006 (a Type
Ia remnant located high above the Galactic plane). The polarization
degree varies in the opposite sense, decreasing with increasing CSM
interaction. It is conceivable that high levels of turbulence are
associated with higher CSM densities, and that this results in a
low polarization degree downstream. Models for the development of
turbulent cascades predict different polarization properties for
different levels of initial isotropy, with isotropic fields yielding
small PD values~\citep{2020Bykov}, similar to those observed for
Cas~A~\citep{Vink+2022}.

\begin{table}[H]
\caption{Polarization
 and physical parameters for \ixpe\ SNRs.\label{tab1}}
		\newcolumntype{C}{>{\centering\arraybackslash}X}
		\begin{tabularx}{\textwidth}{lccccccc}
\toprule
 & \multicolumn{3}{c}{{\textbf{Polarization Degree (\%)~\boldmath{$^\text{a}$}}}} 
& {\boldmath{$V_{shock}$}} & {\boldmath{$n_0$}} &
{\textbf{Bohm}} & {\boldmath{$\rm B_{low}^b$}} \\
 & {\textbf{Rim}} & {\textbf{SNR}} & {\textbf{Peak}} & \textbf{(km s\boldmath{$^{-1}$})} & \textbf{(cm\boldmath{$^{-3})$}} & {\textbf{Factor (\boldmath{$\eta$})}} & \boldmath{($\upmu$\textbf{G})} \\
\midrule

Cas~A & $4.5 \pm 1.0$ & $2.5 \pm 0.5$ & $\sim$15  & $\sim$5800  & $0.9 \pm
0.3$ & $\sim$1--6 & 25--40 \\

Tycho & $12 \pm 2$ & $9 \pm 2$ & $23 \pm 4$ & $\sim$4600 & $\sim$0.1--0.2  &
$\sim$1--5  & 30--40 \\

SN~1006~(NE) & $22.4 \pm 3.5$ & ... & $31 \pm 8$ & $\sim$5000  & $\sim$
0.05--0.08  & $\sim$6--10  & 18--26 \\

RX J1713~(W) & $13.0 \pm 3.5$ & ... & $46 \pm 10$ & 1400--2900 & $\sim$0.01--0.2 & $\sim$1.4  & $\sim$10  \\
\bottomrule
\end{tabularx}

\noindent{\footnotesize{(a) X-ray polarization degree for SNR rim, entire SNR, 
and peak value within SNR. (b) Lower limit to post-shock magnetic field based on
rim width, e.g.,~\citep{VinkLaming2003}.}}
\end{table}

Importantly, for Cas A, Tycho, and SN~1006, the measured polarization
angle implies radial magnetic fields very close to the acceleration
sites. Because particle acceleration in strong non-relativistic
shocks is most efficient for parallel and quasi-parallel magnetic
fields~\citep{CaprioliSpitkovsky2014}, this result may reflect a
selection effect, with emission arising predominantly from regions
where acceleration is efficient~\citep{West+2017}. However, turbulence
driven by rippled shocks resulting from encounters with upstream
density fluctuations can result in dynamo-induced magnetic field
amplification with the radial component dominating the downstream
field, even for largely isotropic upstream turbulence~\citep{Inoue+2013}.
Simulations~\citep{Bykov+2024} show that density fluctuations
associated with the Bell instability~\citep{Bell2004} produced by
streaming cosmic ray currents can produce such an effect, yielding
predominately radial fields in the downstream region under conditions
of high shock velocities and modest density.

For \rxj\, the X-ray polarization results are quite different. While
the polarization degree at the SNR rim is low, as for other SNRs
for which the Bohm factor, $\eta$, approaches 1 (see Table \ref{tab1}), the
inferred magnetic field direction is tangential instead of radial.
This is typical of older SNRs in which the field in the post-shock
region is dominated by the component compressed by the shock. For
fast shocks with strong Bell instabilities, the radial magnetic
field component can peak close to the shock and dominate the
tangential component, while for slower shocks like those in \rxj,
the radial field may peak further downstream, beyond the synchrotron
cooling length of the X-ray-emitting electrons, thus resulting in
a dominant tangential field~\citep{Bykov+2024}.

\ixpe\ studies of additional SNRs, with a range of ages and inferred
Bohm factors, are underway to provide additional constraints on the
connection between polarization properties and the remnant conditions.
Measurements of the polarization degree and magnetic field geometry
will continue to provide new insights into the conditions leading to
efficient particle acceleration in fast shocks.

\vspace{6pt} 

\authorcontributions{
Conceptualization, P.S., R.F., P.Z, and J.V.; investigation, P.S., R.F., P.Z,
and J.V.; writing, review, and editing, P.S., R.F., P.Z, and J.V..
All authors have read and agreed to the published version of the manuscript.
}

\funding{
P.S. acknowledges partial support from NASA Contract NAS8-03060 and NASA
Grant GO1-22057X. P.Z. acknowledges the support from NSFC grant No. 12273010.
R.F. is partially supported by MAECI with grant CN24GR08 ``GRBAXP:
Guangxi-Rome Bilateral Agreement for X-ray Polarimetry in Astrophysics''.
J.V.'s work on IXPE was partially supported by funding from the European Union’s Horizon 2020 research and innovation program under grant agreement No. 101004131 (SHARP).

The Italian contribution to IXPE is supported by the Italian Space Agency (Agenzia Spaziale Italiana, ASI) through contract ASI-OHBI-2022-13-I.0, agreements ASI-INAF-2022-19-HH.0 and ASI-INFN-2017.13-H0, and its Space Science Data Center (SSDC) with agreements ASI-INAF-2022-14-HH.0 and ASI-INFN 2021-43-HH.0, and by the Istituto Nazionale di Astrofisica (INAF) and the Istituto Nazionale di Fisica Nucleare (INFN) in Italy.  

The US contribution to IXPE is supported by the National Aeronautics and Space Administration (NASA) and led and managed by its Marshall Space Flight Center (MSFC), with industry partner Ball Aerospace (contract NNM15AA18C).
}

\conflictsofinterest{
The authors declare no conflict of interest.
}

\begin{adjustwidth}{-\extralength}{0cm}
\reftitle{References}

\PublishersNote{}
\end{adjustwidth}
\end{document}